# STATISTICAL INFERENCE FOR STATISTICAL DECISIONS

Charles F. Manski
Department of Economics and Institute for Policy Research
Northwestern University

September 2019

## Abstract

The Wald development of statistical decision theory addresses decision making with sample data. Wald's concept of a statistical decision function (SDF) embraces all mappings of the form [data → decision]. An SDF need not perform *statistical inference*; that is, it need not use data to draw conclusions about the true state of nature. *Inference-based* SDFs have the sequential form [data → inference → decision]. This paper motivates inference-based SDFs as practical procedures for decision making that may accomplish some of what Wald envisioned. The paper first addresses binary choice problems, where all SDFs may be viewed as hypothesis tests. It next considers *as-if* optimization, which uses a point estimate of the true state as if the estimate were accurate. It then extends this idea to *as-if* maximin and minimax-regret decisions, which use  point estimates of some features of the true state as if they were accurate. The paper primarily uses finite-sample maximum regret to evaluate the performance of inference-based SDFs. To illustrate abstract ideas, it presents specific findings concerning treatment choice and point prediction with sample data.

I am grateful to Ivan Canay, Gary Chamberlain, Kei Hirano, Joel Horowitz, Valentyn Litvin, Bruce Spencer, and Alex Tetenov for comments on drafts of this paper.

1. Introduction

Early in the modern development of econometrics, Haavelmo (1943) differentiated two objectives for empirical research. He wrote (p. 10):

> "The economist may have two different purposes in mind when he constructs a model . . . . First, he may consider himself in the same position as an astronomer; he cannot interfere with the actual course of events. So he sets up the system . . . . as a tentative description of the economy. If he finds that it fits the past, he hopes that it will fit the future. On that basis he wants to make predictions, assuming that no one will interfere with the game. Next, he may consider himself as having the power to change certain aspects of the economy in the future. If then the system . . . . has worked in the past, he may be interested in knowing it as an aid in judging the effect of his intended future planning, because he thinks that certain elements of the old system will remain invariant."

The first research objective is inference and the second is decision making. This paper considers the relationship between the two objectives.

The Wald (1939, 1950) development of statistical decision theory addresses decision making with sample data. Wald began with the standard decision theoretic problem of a planner (equivalently, decision maker or agent) who must choose an action yielding welfare that depends on an unknown state of nature. The planner specifies a state space listing the states that he considers possible. He must choose an action without knowing the true state.

Wald added to this standard problem by supposing that the planner observes sample data that may be informative about the true state. He studied choice of a *statistical decision function (SDF)*, which maps each potential data realization into a feasible action. He proposed evaluation of SDFs as procedures, chosen prior to observation of the data, specifying how a planner would use whatever data may be realized. Thus, Wald's theory is frequentist.

Haavelmo and other early econometricians were concerned with decision making and knew of Wald's work. However, econometrics as a field did not embrace statistical decision theory. Instead, it focused on statistical inference; that is, the use of sample data to draw conclusions about the true state of nature. The broad word "conclusions" has multiple formal interpretations in the literature on inference. To a conditional Bayesian, it means a posterior distribution on the state space. To a frequentist, it means a point or set-valued estimate of the true state. Set-valued estimates include confidence sets and the results of hypothesis tests.

Wald's concept of a statistical decision function embraces all mappings of the form [data → decision]. In general, a SDF need not perform statistical inference. None of the prominent decision criteria that have been studied from Wald's perspective—maximin, minimax-regret (MMR), and maximization of subjective average welfare (minimization of Bayes risk, or simply Bayes)—makes reference to inference. The general absence of inference in statistical decision theory is striking and has been noticed. See Neyman (1962) and Blyth (1970).

Although SDFs need not perform inference, they may do so. That is, they may have the sequential form [data → inference → decision], first performing some form of inference and then using the inference to make a decision. There seems to be no accepted term for such SDFs, so I call them *inference-based*.

A familiar use of inference in decision making is choice between two actions based on the result of a hypothesis test. This has been particularly common when using data from randomized trials to choose between two medical treatments. A common case views one treatment as the status quo and the other as an innovation. The null hypothesis is that the innovation is no better than the status quo and the alternative is that the innovation is better. A considerable body of medical literature recommends that the status quo treatment be used in clinical practice if the null is not rejected and that the innovation be used if it is rejected. Conventional tests fix the probability of Type I and Type II errors at predetermined values, typically 0.05 and 0.20. Manski and Tetenov (2016) show that conventional tests generally do not provide a satisfactory decision theoretic criterion for treatment choice. We recommend non-traditional tests that minimize maximum regret, which symmetrically weights the probabilities and magnitudes of errors.

Another familiar practice uses the sample data to compute a point estimate of the true state of nature, plugs the estimate into the welfare function, and chooses an action that optimizes welfare as if this estimate is accurate. Researchers often justify use of point estimates in decision making by citing limit theorems of asymptotic theory. They observe that an estimate is consistent, asymptotically normal, and so on. However, a planner does not seek an asymptotically attractive estimate of the true state. He seeks to make a good decision with potentially available finite sample data. Citation of favorable asymptotic properties of point estimates does not per se provide a firm foundation for their use in decision making.

A famous finding relating decision making to inference is that decisions using Bayesian inference maximize subjective average welfare. Wald studied maximization of subjective average welfare, also known as minimization of Bayes risk. This decision criterion makes no reference to inference. It simply places a subjective distribution on the state space and optimizes the resulting subjective average welfare. However, examination of the optimization problem shows that the solution is an inference-based SDF when the set of feasible statistical decision functions is unconstrained. The paradigm of *conditional Bayes* decision making calls on one to first perform Bayesian inference, which transforms the prior distribution on the state space into a posterior distribution without reference to a decision problem. One then choose an action that maximizes posterior subjective average welfare. This procedure, applied to each possible data realization, solves Wald's problem of minimization of Bayes risk. See, for example, Berger (1985, Section 4.4.1).

Concern for tractability may motivate interest in inference-based SDFs. In principle, a planner wanting to apply statistical decision theory just needs to specify a decision criterion and apply it to his choice problem. In practice, it may be intractable to determine a Bayes, maximin, or MMR decision. Much of the difficulty stems from the generality of Wald's definition of a statistical decision function, which embraces all mappings of the form [data → decision]. This function space encompasses all possibilities in principle, but the breadth of options can be hard to exploit in practice. When one is unable to apply statistical decision theory in the broad way that Wald envisioned, one may find it expedient to consider relatively simple rules with reasonable decision theoretic properties. Inference-based SDFs may be plausible candidates.

Treatment choice with data from a randomized trial provides an example. Exact computation of the MMR treatment rule is generally complex. It is much simpler to compute the *empirical success* (ES) rule, which chooses the treatment with the highest observed average outcome in the trial. It has been found that the ES rule approximates the MMR rule well in common trial designs. See Manski (2004, 2007a), Hirano and Porter (2009), and Stoye (2009). Kitagawa and Tetenov (2018) extend the ES rule to *empirical welfare maximization* and study some of its regret properties.

Another practical reason for study of inference-based SDF may be necessity. Wald supposed that a planner observes data and makes a decision. In practice, there often is an institutional separation between research and decision making. Researchers observe data and report inferences to the public. Planners do not observe the data, only the reported inferences. Then planners can perform the mapping [inference → decision], but they cannot perform the more basic mapping [data → decision]. They can only choose among the inference-based SDFs made available by research reporting practices.

This paper considers the use of statistical inference in decision making, from the Wald perspective. My aim is not to prove new theorems, but rather to make conceptual points that I feel have not been adequately appreciated. Section 2 reviews the basic concepts of statistical decision theory and discusses the practical difficulty of implementing Wald's broad vision. I then discuss various forms of inference-based SDFs. Section 3 addresses binary choice problems, where all SDFs may be viewed as hypothesis tests. Section 4 considers the familiar practice of *as-if* optimization, where one uses a point estimate of the true state as if the estimate is accurate. Section 5 extends this idea to *as-if* maximin and minimax-regret decisions, which use set estimates of the true state as if they were accurate, in the sense that the set estimate contains the true state. Section 6 concludes.

This paper does not address motivations for statistical inference other than decision making. Researchers sometimes remark that they perform inference in the service of science or knowledge. I do not attempt to interpret this motivation.

## 2. Concepts and Practicalities of Statistical Decision Theory

In this section I first review the basic concepts of statistical decision theory, considering decision making without sample data (Section 2.1) and with data (Sections 2.2). Attention then turns to the practical issues that motivate study of inference-based SDFs (Sections 2.3 through 2.5).

### 2.1. Decisions Under Uncertainty

Consider a planner who must choose an action yielding welfare that varies with the state of nature. The planner has an objective function and beliefs about the true state. These are considered primitives. He must choose an action without knowing the true state.

Formally, the planner faces choice set C and believes that the true state lies in set S, called the state space. The objective function $w(\cdot, \cdot): C \times S \rightarrow R^1$ maps actions and states into welfare. The planner ideally would maximize $w(\cdot, s^*)$, where $s^*$ is the true state. However, he only knows that $s^* \in S$.

The choice set is commonly considered to be predetermined. The welfare function and the state space are subjective. The former formalizes what the decision maker wants to achieve and the latter expresses what states of nature he believes could possibly occur.

A close to universally accepted prescription for decision making is that choice should respect dominance. Action $c \in C$ is weakly dominated if there exists a $d \in C$ such that $w(d, s) \geq w(c, s)$ for all $s \in S$ and $w(d, s) > w(c, s)$ for some $s \in S$. Even though the true state $s^*$ is unknown, choice of d is certain to weakly improve on choice of c.

There is no clearly best way to choose among undominated actions, but decision theorists have not wanted to abandon the idea of optimization. So they have proposed various ways of using the objective function $w(\cdot, \cdot)$ to form functions of actions alone, which can be optimized. In principle one should only consider undominated actions, but it may be difficult to determine which actions are undominated. Hence, it is common to optimize over the full set of feasible actions. I define decision criteria accordingly throughout this paper. I also use max and min notation throughout, without concern for the mathematical subtleties that sometime make it necessary to suffice with sup and inf operations.

One broad idea is to place a subjective distribution on the state space, average state-dependent welfare with respect to this distribution, and maximize the resulting function. This yields maximization of subjective average welfare. Let π be the specified probability distribution on S. For each action c, ∫w(c, s)dπ is the mean of w(c, s) with respect to π. The criterion solves the problem

(1) $\max_{c \,\in\, C} \ \int w(c, s)d\pi.$

Another broad idea is to seek an action that, in some well-defined sense, works uniformly well over all elements of S. This yields the maximin and MMR criteria. The maximin criterion maximizes the minimum welfare attainable across the elements of S. For each feasible action c, consider the minimum feasible value of w(c, s); that is, $\min_{s \,\in\, S} w(c, s)$. A maximin rule chooses an action that solves the problem

(2) $\max_{c \,\in\, C} \ \min_{s \,\in\, S} \ w(c, s).$

The MMR criterion chooses an action that minimizes the maximum loss to welfare that can result from not knowing the true state. A MMR choice solves the problem

(3) $\min_{c \,\in\, C} \ \max_{s \,\in\, S} \ [\max_{d \,\in\, C} w(d, s) - w(c, s)].$

Here $\max_{d \,\in\, C} w(d, s) - w(c, s)$ is the *regret* of action c in state of nature s; that is, the welfare loss associated with choice of c relative to an action that maximizes welfare in state s. The true state being unknown, one evaluates c by its maximum regret over all states and selects an action that minimizes maximum regret. The maximum regret of an action measures its maximum distance from optimality across all states. Hence, a MMR choice is uniformly nearest to optimal among all feasible actions.

A planner who asserts a partial subjective distribution on the states of nature could maximize minimum subjective average welfare or minimize maximum average regret. These hybrid criteria combine elements of averaging across states and concern with uniform performance across states. Hybrid criteria may be of interest. However, I will confine discussion to the polar cases in which the planner asserts a complete subjective distribution or none at all.

## 2.2. Statistical Decision Problems

Statistical decision problems add to the above structure by supposing that the planner observes data drawn from a sampling distribution. In practice, knowledge of the sampling distribution is generally incomplete. To express this, one extends the concept of the state space S to list the set of feasible sampling distributions, denoted $(Q_s, s \in S)$.

Let $\Psi_s$ denote the sample space in state s; that is, $\Psi_s$ is the set of samples that may be drawn under sampling distribution $Q_s$. The literature typically assumes that the sample space does not vary with s and is known. I maintain this assumption and denote the known sample space as $\Psi$, without the s subscript. Then a statistical decision function c(·): $\Psi \rightarrow C$ maps the sample data into a chosen action.

SDF c(·) is a deterministic function after realization of the sample data, but it is a random function ex ante. Hence, the welfare achieved by c(·) is a random variable ex ante. Wald's central idea was to evaluate the performance of c(·) in state s by $Q_s\{w[c(\psi), s]\}$, the ex ante distribution of welfare that it yields across realizations $\psi$ of the sampling process.

It remains to ask how a planner might compare the welfare distributions yielded by different SDFs. The planner wants to maximize welfare, so it seems self-evident that he should prefer SDF d(·) to c(·) in state s if $Q_s\{w[d(\psi), s]\}$ stochastically dominates $Q_s\{w[c(\psi), s]\}$. It is less obvious how he should compare rules whose welfare distributions do not stochastically dominate one another. Wald proposed measurement of the performance of $c(\cdot)$ in state s by its expected welfare across samples; that is, $E_s\{w[c(\psi), s]\} \equiv \int w[c(\psi), s]dQ_s$. An alternative that has drawn only slight attention is to measure performance by quantile welfare (Manski and Tetenov, 2014). Writing in a context where one wants to minimize loss rather than maximize welfare, Wald used the term *risk* to denote the mean performance of a SDF across samples and the term *inadmissible* to denote weak dominance when evaluating performance by risk.

In practice, one does not know the true state. Hence, one evaluates $c(\cdot)$ by the state-dependent expected welfare vector ($E_s\{w[c(\psi), s]\}$, $s \in S$). This done, statistical decision theory can use the same decision criteria as does decision theory without sample data. Let $\Gamma$ be a specified set of feasible SDFs that map $\Psi \to C$. The statistical versions of decision criteria (1), (2), and (3) are

(4)    $\max_{c(\cdot) \in \Gamma} \int E_s\{w[c(\psi), s]\} \, d\pi$,

(5)    $\max_{c(\cdot) \in \Gamma} \min_{s \in S} E_s\{w[c(\psi), s]\}$,

(6)    $\min_{c(\cdot) \in \Gamma} \max_{s \in S} (\max_{d \in C} w(d, s) - E_s\{w[c(\psi), s]\})$.

None of criteria (4) through (6) makes explicit reference to inference. Each directly chooses an SDF that solves the specified problem.

## 2.3. Practicalities

Wald's development of statistical decision theory has breathtaking generality. In principle, it enables comparison of all statistical decision functions whose risk functions exist. Evaluation of performance by quantile welfare eliminates even this requirement. Wald's framework enables comparison of alternative sampling processes as well as decision rules. It applies to any sample size, with no asymptotic approximations. It applies whatever information the decision maker may have. The state space may be finite dimensional or nonparametric. The true state of nature need not be consistently estimable as sample size increases. Thus, the theory is applicable when the true state is a partially identified probability distribution; see Manski (2003) for background on partial identification.

Given these features, one might anticipate that statistical decision theory would play a central role in modern statistics and econometrics. Notable contributions emerged in the 1950s and 1960s, as described in the monographs of Ferguson (1967) and Berger (1985). However, the early period of development of statistical decision theory largely closed by the 1970s, with the exception of the conditional Bayes version of Bayesian theory. Conditional Bayes analysis has continued to develop, but as a self-contained field of study disconnected from Wald's frequentist idea of minimization of Bayes risk.

Why did statistical decision theory lose momentum? One reason may have been diminishing interest in decision making as the motivation for analysis of sample data. Modern statisticians and econometricians tend to view their objective as inference per se, rather than the use of inference in decision making. A different reason may have been the technical difficulty of the subject. Wald's ideas are easy to describe abstractly, but they can be difficult to apply in practice.

Consider the mathematical problems (4) through (6) that define Bayes, maximin, and minimax-regret decisions. These problems are generally well-posed in principle, but they often are not solvable in practice. Each problem requires performance of three nested operations. The inner operation integrates across the sampling distribution of the data to determine expected welfare when a specified SDF is used in a specified state of nature. The middle operation integrates or finds an extremum of the result of the inner operation across the state space. The outer operation finds an extremum of the result of the middle operation across all SDFs.

Analytical arguments and numerical computation occasionally yield tractable solutions to these problems. Early analytical work in the conditional Bayes paradigm studied *conjugate priors*, which pair certain prior distributions on the state space with certain state-dependent sampling distributions for the data to yield simple posterior distributions. Analytical solutions to certain MMR problems include the early study by Hodges and Lehmann (1950) of point prediction with data from a random sample and the recent study by Stoye (2009) of treatment choice with data from a randomized experiment.

Numerical computation sometimes provides a practical alternative to analysis. Numerical solution was generally not feasible when statistical decision theory originated in the middle of the twentieth century, but it has become increasingly possible since then. Modern conditional Bayes analysis has increasingly moved away from use of conjugate priors to numerical computation of posterior distributions. Numerical determination of some maximin and MMR decisions has also become feasible. For example, Manski (2007a, Chapter 12) and Manski and Tetenov (2016, 2019) present in tabular form the MMR solutions to certain treatment choice problems with data from a randomized experiment. Computation of state-dependent expected welfare, the inner operation in problems (4) through (6), can now be accomplished numerically by Monte Carlo integration methods. Manski and Tabord-Meehan (2017) give an application that will be discussed in Section 5.

Future advances in analysis and numerical computation should continue to expand the settings in which it is tractable to determine Bayes, maximin, and MMR decisions. However, I doubt that it will soon become possible to implement these decision criteria in generality. I do not foresee imminent discovery of a practical general way to carry out the three nested operations that define problems (4) through (6).

The absence of a practical way to implement Wald's broad vision motivates study of inference-based SDFs. The prevalent practice in research on inference per se, divorced from decision making, has been to evaluate the performance of hypothesis tests by their power and that of point estimates by their limit properties as sample size grows. Statistical decision theory recommends that, when the objective is to make a decision, the performance of tests and estimates should be evaluated by the expected welfare that they yield.

Recall that Wald's central idea is to evaluate the performance of SDF $c(\cdot)$ in any state s by $Q\{w[c(\psi), s]\}$, the distribution of welfare that it yields across realizations $\psi$ of the sampling process. This central idea may be applied when a inference-based SDF is used to make a decision. One may measure the performance of a hypothesis test or a point estimate by the resulting value of subjective average welfare, minimum expected welfare, or maximum regret. Computation of these quantities requires carrying out the inner and middle operations of problems (4) through (6), but not the outer operation.

2.4. Focus on Maximum Regret

In the remainder of this paper, I measure the performance of SDFs by maximum regret rather than by maximum subjective average welfare or minimum expected welfare. This section explains why.

I have no objection to computation of subjective average welfare when one feels able to place a credible subjective prior distribution on the state space. However, Bayesians have long struggled to provide guidance on specification of priors and the matter continues to be controversial. The controversy suggests that inability to express a credible prior is common in actual decision settings.

When one finds it difficult to assert a credible subjective distribution, Bayesians who believe it essential to use a probability distribution to express uncertainty may suggest use of some default distribution that is variously called a "reference" or "conventional" or "objective" prior; see, for example, Berger (2006). However, there is no consensus on the prior that should play this role. The chosen prior matters for decision making.

In the absence of a credible prior distribution on the state space, practical and conceptual reasons motivate measurement of the performance of SDFs by maximum regret, rather than by minimum expected welfare. From a practical perspective, it has been found that MMR decisions behave more reasonably than do maximin ones in the important context of treatment choice. In common settings of treatment choice with trial data, it has been found that the MMR rule is well approximated by the ES rule, which chooses the treatment with the highest observed average outcome in the trial. In contrast, the maximin criterion commonly ignores the trial data, whatever they may be. This was recognized verbally by Savage (1951), who stated that the criterion is "ultrapessimistic" and wrote (p. 63): "it can lead to the absurd conclusion in some cases that no amount of relevant experimentation should deter the actor from behaving as though he were in complete ignorance." Savage did not flesh out this statement, but it is easy to show that this occurs with trial data. Manski (2004) provides a simple example.

The conceptual appeal of using maximum regret to measure performance is that maximum regret quantifies how lack of knowledge of the true state of nature diminishes the quality of decisions. While the term "maximum regret" has become standard in the literature, this term is a shorthand for the maximum sub-optimality of a decision criterion across the feasible states of nature. An STR with small maximum regret is uniformly near-optimal across all states. This is a desirable property.

## 2.5. Decision Making with Models of the State Space

To close this opening discussion of statistical decision theory, I remark on decision making with models of the state space.

I stated at the outset that standard decision theory begins with a planner who "specifies a state space listing the states that he considers possible." Thus, the state space should include all states that the planner believes feasible. The terms "considers possible" and "believes feasible" are necessarily subjective and context-specific.

The state space may be a large set that is difficult to contemplate in its entirety. Hence, it is common to perform inference and make decisions using a model of the state space. The word "model" is commonly used informally to connote a simplification or approximation of reality. Formally, a model specifies an alternative to the state space. Thus, model m replaces S with a model space $S_m$. A planner using a model to make a decision acts as if the model space is the state space. Thus, the planner might solve problem (4), (5), or (6) with $S_m$ replacing S. A decision using a model need not be inference-based. However, researchers often perform inference with models and planners often use these inferences to make decisions.

The states contained in a model space may or may not be elements of the state space. The statistician George Box famously wrote "All models are wrong but some are useful" (Box, 1979). The phrase "all models are wrong" indicates that Box was thinking of models that simplify or approximate reality in a way that one believes could not possibly be accurate; then $S_m \cap S = \varnothing$. Often, however, researchers report inferences on models that they believe could possibly be accurate but are not necessarily so; then $S_m \subset S$.

However one specifies a model and uses it to make a decision, statistical decision theory provides a clear way to evaluate performance. What matters is the SDF resulting from use of a model. As with any SDF, the performance of one using model m may be measured by its state-dependent expected welfare. The relevant states for evaluation of performance are those in the state space S, not those in the model space $S_m$. In particular, maximum regret (uniform nearness to optimality) provides an informative scalar measure of performance.

Thus, the Wald theory enables one to operationalize Box's assertion that some models "are useful." Useful models are ones whose use in decision making yields acceptably high state-dependent expected welfare, relative to what is possible in principle.

3. Binary Choice and Hypothesis Tests

SDFs for binary choice problems can always be viewed as hypothesis tests. To see this, let choice set C contain two actions, say C = {a, b}. A SDF c(·) partitions $\Psi$ into two regions that separate the data yielding choice of each action. These regions are $\Psi_{c(\cdot)a} \equiv [\psi \in \Psi: c(\psi) = a]$ and $\Psi_{c(\cdot)b} \equiv [\psi \in \Psi: c(\psi) = b]$.

A hypothesis test motivated by the choice problem partitions state space S into two regions, say $S_a$ and $S_b$, that separate the states in which actions a and b are uniquely optimal. Thus, $S_a$ contains the states [s $\in$ S: w(a, s) > w(b, s)] and $S_b$ contains [s $\in$ S: w(b, s) > w(a, s)]. The choice problem does not provide a rationale for allocation of states in which the two actions yield equal welfare. The standard practice in testing is to give one action, say a, a privileged status and to place all states yielding equal welfare in $S_a$. Then $S_a \equiv [s \in S: w(a, s) \geq w(b, s)]$ and $S_b \equiv [s \in S: w(b, s) > w(a, s)]$.

In the language of hypothesis testing, SDF $c(\cdot)$ performs a test with acceptance regions $\Psi_{c(\cdot)a}$ and $\Psi_{c(\cdot)b}$. When $\psi \in \Psi_{c(\cdot)a}$, $c(\cdot)$ accepts the hypothesis $\{s \in S_a\}$ by setting $c(\psi) = a$. When $\psi \in \Psi_{c(\cdot)b}$, $c(\cdot)$ accepts the hypothesis $\{s \in S_b\}$ by setting $c(\psi) = b$. I use the word "accepts" rather than the traditional term "does not reject" because choice of a or b is an affirmative action.

Although all SDFs for binary choice are tests, classical hypothesis testing and statistical decision theory evaluate tests in fundamentally different ways. Sections 3.1 and 3.2 contrast the two paradigms in general terms. Sections 3.3 and 3.4 give specific examples.

## 3.1. Classical Tests

Let us review the basic practices of classical hypothesis testing, as developed by Neyman and Pearson (1928, 1933). Classical tests view the two hypotheses $\{s \in S_a\}$ and $\{s \in S_b\}$ asymmetrically, calling the former the null hypothesis and the latter the alternative. The sampling probability of rejecting the null hypothesis when it is correct is the probability of a Type I error. A longstanding convention has been to restrict attention to tests in which the probability of a Type I error is no larger than some predetermined value $\alpha$, usually 0.05, for all $s \in S_a$. In the notation of statistical decision theory, one restricts attentions to SDFs $c(\cdot)$ for which $Q_s[c(\psi) = b] \leq \alpha$ for all $s \in S_a$.

Among tests that satisfy this restriction, classical testing seeks ones that give small probability of rejecting the alternative hypothesis when it is correct, the probability of a Type II error. However, it generally is not possible to attain small probability of a Type II error for all $s \in S_b$. Letting S be a metric space, the probability of a Type II error typically approaches $1 - \alpha$ as $s \in S_b$ nears the boundary of $S_a$. See, for example, Manski and Tetenov (2016), Figure 1. Given this, the convention has been to restrict attention to states in $S_b$ that lie at least a specified distance from $S_a$.

Let $\rho$ be the metric measuring distance on S. Let $\rho_a > 0$ be the specified minimum distance from $S_a$. In the notation of statistical decision theory, classical testing seeks small values for the maximum value of $Q_s[c(\psi) = a]$ over $s \in S_b$ s.t. $\rho(s, S_a) \geq \rho_a$.

## 3.2. Decision Theoretic Evaluation of Tests

Decision theoretic evaluation of tests does not restrict attention to tests that yield a predetermined upper bound on the probability of a Type I error. Nor does it aim to minimize a constrained maximum value of the probability of a Type II error. Wald's central idea for binary choice as elsewhere is to evaluate the performance of SDF $c(\cdot)$ in state s by the distribution of welfare that it yields across realizations of the sampling process. He first addressed hypothesis testing this way in Wald (1939).

The welfare distribution in a binary choice problem is Bernoulli, with mass points max [w(a, s), w(b, s)] and min [w(a, s), w(b, s)]. These mass points coincide if w(a, s) = w(b, s). When s is a state where w(a, s) ≠ w(b, s), let $R_{c(\cdot)s}$ denote the probability that $c(\cdot)$ yields an error, choosing the inferior treatment over the superior one. That is,

(7)     $R_{c(\cdot)s} = Q_s[c(\psi) = b]$   if w(a, s) > w(b, s),

         $= Q_s[c(\psi) = a]$   if w(b, s) > w(a, s).

The former and latter are the probabilities of Type I and Type II errors. Whereas classical testing treats these error probabilities differently, statistical decision theory views them symmetrically.

The probabilities that welfare equals max [w(a, s), w(b, s)] and min [w(a, s), w(b, s)] are $1 - R_{c(\cdot)s}$ and $R_{c(\cdot)s}$. Wald measured the performance of SDFs by expected welfare. In binary choice problems, expected welfare in state s is

(8)    $E_s\{w[c(\psi), s]\} = R_{c(\cdot)s}\{\min [w(a, s), w(b, s)]\} + [1 - R_{c(\cdot)s}]\{\max [w(a, s), w(b, s)]\}$

$$= \max [w(a, s), w(b, s)] - R_{c(\cdot)s}\cdot|w(a, s) - w(b, s)|.$$

The expression $R_{c(\cdot)s}\cdot|w(a, s) - w(b, s)|$ is the regret of $c(\cdot)$ in state s. Thus, regret is the product of the error probability and the magnitude of the welfare loss when an error occurs.

Computation of regret in a specified state is usually practical. The welfare magnitudes $w(a, s)$ and $w(b, s)$ are usually easy to compute. The error probability $R_{c(\cdot)s}$ typically does not have a simple explicit form, but it usually can be approximated to any desired precision by Monte Carlo integration. That is, one draws repeated pseudo-realizations of $\psi$ from the distribution $Q_s$, computes the fraction of cases in which the resulting $c(\psi)$ selects the inferior action, and uses this to estimate $R_{c(\cdot)s}$.

Whereas computation of regret in one state is not problematic, computation of maximum regret across all feasible states may be burdensome. The state space commonly is uncountable in applications. A standard practice when considering problems with uncountable state spaces is to discretize the space, limiting attention to a finite subset of states that reasonably approximate the full state space. Section 3.4 illustrates.

## 3.3. A Medical Example

Manski (2019) gives a simple hypothetical medical example illustrating how classical and decision theoretic evaluation of tests differ. I paraphrase here.

Suppose that a typically terminal form of cancer may be treated by a status quo treatment or an innovation. It is known from experience that mean patient life span with the status quo treatment is one year. Prior to use of the innovation, medical researchers see two possibilities for its effectiveness. It may be less effective than the status quo, yielding a mean life span of only 1/3 of a year, or it may be much more effective, yielding a mean life span of 5 years.

A randomized trial is performed to learn the effectiveness of the innovation. The trial data are used to perform a conventional test comparing the innovation and the status quo. The null hypothesis is that the innovation is no more effective than the status quo and the alternative is that the innovation is more effective. The probabilities of Type I and Type II errors are 0.05 and 0.20. The test result is used to choose between the treatments.

If the status quo treatment is superior, a Type I error occurs with sampling probability 0.05 and reduces mean patient life span by 2/3 of a year (1 year minus 1/3 year), so regret is 1/30 of a year. If the innovation is superior, a Type II error occurs with probability 0.20 and reduces mean patient life span by 4 years (5 years minus 1 year), so regret is 4/5 of a year. Use of the test to choose between the status quo and the innovation implies that society is willing to tolerate a large (0.20) chance of a large welfare loss (4 years) when making a Type II error, but only a small (0.05) chance of a small welfare loss (2/3 of a year) when making a Type I error. The theory of hypothesis testing does not motivate this asymmetry.

The maximum regret of the conventional test is 4/5 of a year. Rather than use this test to choose treatment, one could perform a test that has smaller maximum regret. A simple option may be to reverse the conventional probabilities of Type I and Type II errors; thus, one might perform a test with a 0.20 chance of a Type I error and a 0.05 chance of a Type II error. If the status quo treatment is superior, the regret of this unconventional test is 2/15 of a year; that is, a 0.20 chance of a Type I error times a 2/3 of a year reduction in mean life span with improper choice of the innovation. If the innovation is superior, regret is 1/5 of a year; that is, a 0.05 chance of a Type II error times a 4-year reduction in mean life span with improper choice of the status quo. Thus, the maximum regret of the unconventional test is 1/5 of a year.

In this example, the unconventional test delivers much smaller maximum regret than does the conventional test. Other tests may perform even better.

### 3.4. Comparing a Classical Test with One that Minimizes Maximum Regret

Manski and Tetenov (2016) compare a classical test with one that minimizes maximum regret, in a simple context where the MMR decision is known. The context is choice between two treatments, say t = a
and t = b, when the outcome of interest is binary, with y(t) = 1 denoting success and y(t) = 0 failure. State
s indicates a possible value for the pair of outcome probabilities {$P_s$[y(a) = 1], $P_s$[y(b) = 1]}. The welfare
yielded by treatment t in state s is w(t, s) = $P_s$[y(t) = 1].

The sample data are the findings of a balanced randomized trial, assigning the same number N of
subjects to each treatment. In this setting, Stoye (2009) shows that the ES rule is the MMR decision. A widely
used classical test is a one-sided two-sample z-test, which asymptotically makes the probability of a Type I
error equal to 0.05. See Fleiss (1973) for details.

The regret of any test c(·) in any state s is $R_{c(·)s}$·|$P_s$[y(a) = 1] − $P_s$[y(b) = 1]|. The quantity |$P_s$[y(a) =
1] − $P_s$[y(b) = 1]| is the absolute value in state s of the average treatment effect comparing welfare with
treatments a and b. Thus, the regret of a test in state s is the product of its error probability and the absolute
value of the effect size.

We suppose that the planner has no a priori knowledge of the outcome probabilities. Hence, the state
space is the rectangle [0, 1]². We approximate maximum regret by computing regret over a dense finite grid
of states, thus discretizing the state space.

Figure 1 of Manski and Tetenov (2016) shows how the regret incurred by the ES rule and the z-test
rule varies with the effect size for a sample size of N = 145 per treatment arm. Maximum regret is 0.01 for
the ES rule and 0.05 for the z-test. Maximum regret for each test occurs at an intermediate effect size. Regret
is necessarily small for small effect sizes. Regret is also small for large effect sizes, because the probability
of error declines with the effect size. The intermediate effect sizes at which regret is maximized differ for the
two tests, reflecting the differences in their state-specific error probabilities. For the ES rule, regret is
maximized when the effect size is approximately ±0.03 and the error probability is approximately 0.35. For
the z-test, regret is maximized when the effect size is approximately 0.08 and the error probability is
approximately 0.6.

4. As-If Optimization

The Introduction mentioned the common practice of using sample data to compute a point estimate of the true state of nature and choice of an action that optimizes welfare as if this estimate is accurate. This section elaborates. Section 4.1 considers the practice in abstraction. Section 4.2 specializes to the important class of statistical decision problems in which states are probability distributions and the data are a random sample from the true distribution. Sections 4.3 and 4.4 consider use of the empirical success rule from the decision theoretic perspective of maximum regret.

4.1. General Considerations

A point estimate is a function $s(\cdot)$: $\Psi \to S$ that maps data into a state of nature. As-if optimization means solution of the problem $\max_{c \in C} w[c, s(\psi)]$. When as-if optimization yields multiple solutions, one may use some auxiliary rule to select one. Then one obtains the SDF $c[s(\cdot)]$, where

$$(9) \quad c[s(\psi)] \; \equiv \; \underset{c \in C}{\text{argmax}} \; w[c, s(\psi)], \quad \psi \in \Psi.$$

Solving problem (10) is often much simpler than solving the general problems (4) through (6).

Researchers often motivate use of point estimates in decision making by citing limit theorems of asymptotic theory. They hypothesize a sequence of sampling processes indexed by sample size N and a corresponding sequence of point estimates $s_N(\cdot)$: $\Psi_N \to S$. They show that the sequence is consistent when specified assumptions hold. That is, $s_N(\psi) \to s^*$ as $N \to \infty$, in probability or almost surely.

The concept of regret provides a decision theoretic interpretation of the concept of consistency. We may say that a sequence of SDFs $c[s_N(\psi)]$ generated by as-if optimization is pointwise consistent if regret converges to zero in all states of nature as $N \to \infty$. It is uniformly consistent if maximum regret converges to zero. Going beyond consistency, the theoretical statistics literature on *limits of experiments* (Le Cam, 1972) has developed a delicate form of local asymptotic analysis that seeks to provide guidance for decision making. See Hirano and Porter (2009, 2019) for applications to treatment choice.

Asymptotic arguments can be suggestive, but they do not prove that as-if optimization has desirable properties when used with finite samples. As with other SDFs, statistical decision theory evaluates as-if optimization in state s by the welfare sampling distribution $Q_s\{w\{c[s(\psi)], s\}\}$ that it yields. As with other SDFs, Wald would measure performance by expected welfare $E_s\{w\{c[s(\psi)], s\}\}$. Statistical decision theory calls for study of the behavior of expected welfare across the state space.

## 4.2. Solving Sample Analogs of Decision Problems where States are Probability Distributions

Wald's broad vision of statistical decision theory places no restrictions on the nature of the state space and the sample data. Nevertheless, research using the theory has focused to a considerable degree on a class of problems that connect the state space and the sampling distribution in a simple way. These are problems in which states are probability distributions and the data are a random sample drawn from the true distribution.

In such problems, a natural form of as-if optimization is to act as if the empirical distribution of the data is the true population distribution. The usual justification is asymptotic. The empirical distribution consistently estimates the population distribution and has various further desirable asymptotic properties. This suggests use of the empirical distribution in decision making.

Familiar examples of this class of as-if optimization methods occur when states are distributions for a real random variable and the decision problem is to predict the value of a realization drawn at random from the true distribution. When welfare is measured by state-dependent square and absolute loss, the ideal best predictors are the population mean and median. When the true distribution is not known but data from a random sample are observed, as-if optimization suggests use of the sample average and median as predictors.

Another familiar example occurs when states are multivariate distributions of treatment response and the decision problem is to choose treatments for members of the population. When welfare is measured by the mean outcome of treatment across the population, the ideal best treatment maximizes this mean outcome. When the true distribution of treatment response is not known but outcomes in a randomized trial are observed, as-if optimization suggests use of the empirical success rule.

Manski and Tetenov (2016) derive finite-sample bounds on maximum regret when the ES rule is used to choose treatments with data from a randomized trial. Section 4.3 summarizes this analysis.

## 4.3. Bounds on Maximum Regret for Treatment Choice with the Empirical Success Rule

I discussed earlier the MMR property of the ES rule in problems with two treatments and balanced designs. Research to date has not been able to fully characterize the MMR rule in problems with multiple treatments or unbalanced designs. Nevertheless, Manski and Tetenov (2016) use large deviations inequalities for sample averages of bounded outcomes to obtain informative and easily computable upper bounds on the maximum regret of the ES rule. Their Proposition 1 extends an early finding of Manski (2004) from two to multiple treatments. Proposition 2 derives a new large-deviations bound for multiple treatments.

Let K be the number of treatment arms and let M be the range of the bounded outcome. When the trial has a balanced design, with n subjects per treatment arm, the bounds on maximum regret proved in Propositions 1 and 2 have particularly simple forms, being

(10) $\quad (2e)^{-1/2} M(K-1) n^{-1/2},$

(11) $\quad M (\ln K)^{1/2} n^{-1/2}.$

Result (10) provides a tighter bound than (11) for two or three treatments. Result (11) gives a tighter bound for four or more treatments.

## 4.4. Using the Empirical Success Rule when Random Treatment Selection is a Model

The above consideration of the ES rule assumed treatment choice with data from a randomized trial. Treatment selection is random in an ideal trial. Thus, the state space only includes states in which potential treatment response is statistically independent of the treatments that persons in the study population actually receive.

A planner may use the ES rule with data from an observational study or a trial with imperfect compliance. In these settings, random treatment selection may be just a model as discussed in Section 2.6, not a credible assumption. Then the bounds on maximum regret given in (10) and (11) do not hold. Indeed, maximum regret need not converge to zero as sample size increases.

Consider the case with two treatments, say $\{a, b\}$. Each member of the study population has potential outcomes $[y(a), y(b)]$. Let binary indicators $[\delta(a), \delta(b)]$ denote whether these outcomes are observable. In a study where all realized outcomes are observed, only counterfactual outcomes are unobserved. Hence, the possible values for the observability indicators are $[\delta(a) = 1, \delta(b) = 0]$ and $[\delta(a) = 0, \delta(b) = 1]$. In a study with missing data on some realized outcomes, $[\delta(a) = \delta(b) = 0]$ for some persons.

Each element s of the state space denotes a possible distribution $P_s[y(a), y(b), \delta(a), \delta(b)]$ of outcomes and observability. In an ideal trial, $y(\cdot)$ is statistically independent of $\delta(\cdot)$ in every feasible state. This implies that $E_s[y(t)|\delta(t) = 1] = E_s[y(t)]$, $t \in \{a, b\}$ for all $s \in S$. However, this need not hold when the data are not generated by an ideal trial.

Section 3 observed that the regret of any SDF $c(\cdot)$ in state s is $R_{c(\cdot)s} \cdot |E_s[y(b)] - E_s[y(a)]|$. Regret is zero in states where $E_s[y(b)] = E_s[y(a)]$, so it suffices to consider ones where $E_s[y(b)] \neq E_s[y(a)]$. Let $S_+$ denote the states in which $E_s[y(a)|\delta(a) =1] \neq E_s[y(b)|\delta(b) = 1]$, with the same ordering as $E_s[y(a)]$ and $E_s[y(b)]$. Let $S_-$ denote the states in which $E_s[y(a)|\delta(a) = 1] \neq E_s[y(b)|\delta(b) = 1]$, with the reverse order of $E_s[y(a)]$ and $E_s[y(b)]$.

Let $S_=$ denote the states in which $E_s[y(a)|\delta(a) = 1] = E_s[y(b)|\delta(b) = 1]$.

Suppose that $P_s[\delta(a) = 1] > 0$ and $P_s[\delta(b) = 1] > 0$ in all feasible states. Then the ES rule is generically well-defined. As sample size increases, empirical success for treatments a and b converge to $E_s[y(a)|\delta(a) = 1]$ and $E_s[y(b)|\delta(b) = 1]$ respectively. Hence the error probability with the ES rule converges to zero in each of the states in $S_+$ and to one in each of the states in $S_-$. The error probability may not converge in $s \in S_=$. Hence, the maximum regret of the ES rule asymptotically lies in the interval with lower bound $\max_{s \in S_-} |E_s[y(b)] - E_s[y(a)]|$ and upper bound $\max_{s \in S_- \cup S_=} |E_s[y(b)] - E_s[y(a)]|$ .

Recall Box's statement that a model may be wrong but useful. The above derivation indicates that, from the perspective of treatmentchoice, the model of random treatment selection is useful when the sets $S_-$ and $S_=$ are empty. Then the maximum regret of the ES rule converges to zero even though treatment selection may not be random. One might continue to consider this model useful if $S_-$ and $S_=$ are not empty, but $\max_{s \in S_- \cup S_=} |E_s[y(b)] - E_s[y(a)]|$ is small.

## 5. As-If Decisions with Set Estimates of the True State

As-if optimization uses data to compute a point estimate of the true state of nature and chooses an action that optimizes welfare as if this estimate is accurate. An obvious extension is to use data to compute a set-valued estimate of the true state and then act as if the set estimate is accurate. Whereas a point estimate $s(\cdot): \Psi \to S$ maps data into an element of S, a set estimate $S(\cdot): \Psi \to 2^S$ maps data into a subset of S. Given data $\psi$, one acts as if the state space is $S(\psi)$ rather than the larger set S.

In particular, one could solve these data-dependent versions of problems (1) through (3):

(1′)     $\max_{c \in C} \int w(c, s) d\pi(\psi)$,

(2′)     max    min    w(c, s),
         c ∈ C   s ∈ S(ψ)

(3′)     min    max    [max w(d, s) − w(c, s)].
         c ∈ C   s ∈ S(ψ)   d ∈ C

In the case of (1′), $\pi(\psi)$ is a subjective distribution on the set S($\psi$).

From a computational perspective, these as-if problems are generally easier to solve than are the actual maximin and MMR problems with sample data, stated in (4) through (6). The as-if problems fix $\psi$ and select one action c, whereas the actual problems require one to consider all potential samples and choose a decision function c(·). The as-if problems compute welfare values w(c, s), whereas the actual problems must compute more complex expected welfare values $E_s\{w[c(\psi), s]\}$. Sections 5.2 and 5.3 provide examples.

An alternative type of as-if approach replaces S by S($\psi$) in the middle operations of (4) through (6), but it does not replace $E_s\{w[c(\psi), s]\}$ by w(c, s) in the innermost part of each criterion. This approach simplifies (4) through (6) by shrinking the state space over which the middle operations are performed. However, it is more complex than (1′) through (3′) for two reasons. It requires choice of a decision function c(·) rather than a single action c, and it must compute $E_s\{w[c(\psi), s]\}$ rather than w(c, s). Chamberlain (2000) uses asymptotic considerations to suggest this type of as-if decision making and presents an application.

Asymptotic arguments suggest that as-if maximin and MMR decision making should be attractive in settings where states of nature are probability distributions and the sampling process generating data partially identifies the true distribution. Let $H(s^*)$ denote the identification region for the true state $s^*$; that is, the subset of S that the sampling process would ideally reveal with observable population rather than sample data.

Research on inference on partially identified distributions shows that, given modest regularity conditions, the sampling process enables consistent estimation of $H(s^*)$; that is, there exist sequences of set estimates such that $S_N(\psi) \to H(s^*)$ as $N \to \infty$, in probability or almost surely. See Molinari (2019) for review of the literature on inference. Given regularity conditions on the welfare function and choice set, consistency of $S_N(\cdot)$ implies consistency of as-if decision making.

Asymptotic arguments per se suggest, but do not prove, that as-if decision making has desirable finite-sample properties. Sections 5.2 and 5.3 summarize work on two settings with missing data in which as-if MMR decisions have been studied. Section 5.4 briefly considers decision making using confidence sets as set estimates.

## 5.2. As-If MMR Point Prediction with Missing Data

Among the earliest statistical decision problems that have been studied is best point prediction of a real outcome under square loss. The ideal, but unknown, best predictor in this familiar setting is the population mean outcome. The risk of a candidate predictor using finite sample data is the sum of the population variance of the outcome and the mean square error (MSE) of the predictor as an estimate of the mean outcome. The regret of a predictor is its MSE as an estimate of the mean. An MMR predictor minimizes maximum mean square error. Thus, MMR prediction of the outcome is equivalent to minimax estimation of the population mean. I summarize here recent research on MMR prediction when some sample data are missing.

### 5.2.1. Analytical Findings

Hodges and Lehman (1950) derived the MMR predictor with data from a random sample when the outcome has known bounded support and all sample data are observed. Normalizing the support to be the interval [0, 1], Theorem 6.1 proves that the MMR predictor is $(m\sqrt{N} + \frac{1}{2})(\sqrt{N} + 1)^{-1}$, where N is sample size and m is the sample average outcome. One might interpret this as an inference-based SDF, using m to estimate the population mean outcome. However, the MMR predictor does not perform as-if optimization. Rather than consider m to be an accurate estimate of the population mean, it recognizes statistical imprecision by shrinking m toward the value ½.

Dominitz and Manski (2017) study best prediction under square loss with data from a random sample with some missing outcome data. It is computationally challenging to determine the MMR predictor when some data are missing. Seeking an approach that is both tractable and reasonable, the paper studies as-if MMR prediction. The analysis assumes knowledge of the fraction of the population with missing data, but it makes no assumption about the population distribution of missing outcomes.

Lacking knowledge of the distribution of missing outcomes, the population mean outcome is partially identified when the outcome of interest has bounded support. Its identification region is an easy-to-compute interval derived in Manski (1989). If this interval were known, the MMR predictor would be its midpoint. The identification interval is not known with sample data, but one can compute its sample analog and use the midpoint of the sample-analog interval as the predictor.

This *midpoint predictor* is easy to compute. Its maximum regret has a simple analytical form. Let $\delta$ indicate the observability of an outcome. Let $P(\delta = 1)$ and $P(\delta = 0)$ denote the fractions of the population whose outcomes are and are not observable. Let N be the number of observed sample outcomes, which is fixed rather than random under the assumed survey design. The paper proves that the maximum regret of the midpoint predictor is $\frac{1}{4}[P(\delta = 1)^2/N + P(\delta = 0)^2]$.

Dominitz and Manski (2019) build on the work in Dominitz and Manski (2017) to study MMR prediction under square loss of functions of two variables, when some data on one variable are missing. This prediction problem may arise in longitudinal data collection with attrition, if there is a 100-percent response rate in period 1 and some nonresponse in period 2. It may also arise in cross-sectional collection of data on two household members or in surveys with item nonresponse.

Predicting the value of functions of two variables is more complex than functions of one variable, because the form of the function matters. The paper focuses on prediction of linear and indicator functions. It is easy to compute midpoint predictors akin to that posed in the previous study. The paper obtains an analytical expression for maximum regret when predicting the value of an indicator function. It derives an analytical upper bound on maximum regret when predicting the value of linear functions.

5.2.2. Numerical Computation of Maximum Regret

The analytical findings on the maximum regret of midpoint predictors described above assume knowledge of the fraction of the population with missing data. Midpoint predictors remain easy to compute when this fraction is not known and instead is estimated by its sample analog. In this case, derivation of an analytical expression for maximum regret does not seem possible but numerical computation is tractable.

Manski and Tabord-Meehan (2017) describe an algorithm coded in STATA for numerical computation of the maximum regret of the midpoint predictor and other user-specified predictors in the setting of Dominitz and Manski (2017), where the objective is to predict a bounded real outcome. The program, named *wald_mse*, does not require knowledge of the population fraction of missing data. Instead, $P(z = 0)$ may be estimated by its sample analog.

Letting y denote the outcome of interest, the state space has the form $[P_s(y|\delta = 1), P_s(y|\delta = 0), P_s(\delta = 0)]$, $s \in S$. An important feature of *wald_mse* is that the user can specify the state space flexibly. For example, the user may assume that nonresponse will be no higher than 80% or that the mean value of the outcome for nonresponders will be no lower than 0.5. The user may impose no restrictions connecting the two outcome distributions $P_s(y|\delta = 1)$ and $P_s(y|\delta = 0)$, or he may bound the difference between these distributions.

Given any state s, the algorithm uses Monte Carlo integration to approximate the MSE of a user-specified predictor. The quality of the approximation is controlled by user specification of the number of pseudo realizations of $(y, \delta)$ that are drawn. Increasing the number yields a better approximation at the cost of longer computation time.

The algorithm embodies two approaches to maximize MSE across the state space, one applicable when the outcome is binary and the other when the outcome has a continuous distribution. When y is binary, $P_s(y|\delta = 1)$, $P_s(y|\delta = 0)$, and $P_s(\delta = 0)$ are all Bernoulli distributions. The algorithm approximates the state space by a finite grid over the possible Bernoulli parameters for each distribution. It then maximizes MSE over the grid. The user controls the quality of the approximation to the state space by specifying the density of the grid. Increasing the density yields a better approximation at the cost of longer computation time.

When y is continuous, the algorithm presumes that $P_s(y|\delta = 1)$ and $P_s(y|\delta = 0)$ are Beta distributions, while $P_s(\delta = 0)$ is a Bernoulli distribution. Supposing that the two outcome distributions are Beta is a substantive restriction, imposed to achieve a relatively flexible and tractable approximation. The state space is approximated by a finite grid over the possible shape parameters of the Beta distributions and over the possible values of the Bernoulli parameter. As before, the user specifies the density of the grid and the algorithm maximizes mean square error over the grid.

## 5.3. As-If MMR Treatment Choice with Missing Outcome Data

Sections 3 and 4 discussed treatment choice with data from ideal randomized trials, where the only inferential issue is finite-sample statistical imprecision. When choosing treatments with data from non-ideal trials and observational studies, planners must also contend with serious identification problems. Section 4.4 discussed using the ES rule when random treatment selection is a model. However, this approach acts as if identification problems are not present. In general, the ES rule does not minimize maximum regret when identification problems are recognized.

Manski (2007b) studies MMR choice between two treatments {a, b} when some realized outcomes in a study population are unobservable and the distribution of missing data is unknown. The data may be an observational study where one observes a random sample of persons whose treatments were selected in an unknown way; then the planner neither observes the outcomes of non-selected treatments nor knows the distribution of such outcomes. Or the data may be the findings of a randomized trial in which some subjects drop out for unknown reasons before their outcomes can be measured. In these settings, the inferential problems include both partial identification and finite-sample imprecision. The analysis is performed conditional on observed covariates of population members, which I suppress for simplicity here.

As in Section 4.4, each member of the study population has potential outcomes [y(a), y(b)]. Binary indicators [δ(a), δ(b)] again denote whether these outcomes are observable. Each element s of the state space denotes a possible distribution $P_s[y(a), y(b), δ(a), δ(b)]$ of outcomes and observability.

The set C of feasible actions are fractional treatment allocations, assigning the fraction $z(b) \in [0, 1]$ of persons to treatment b and the fraction $z(a) = 1 - z(b)$ to treatment a. Supposing that the population is large and that its members are observationally identical, prospective treatment assignment is necessarily random. The welfare yielded by allocation $z(\cdot)$ in state s is $w[z(\cdot), s] = [1 - z(a)] \cdot E_s[y(a)] + z(b) \cdot E_s[y(b)]$.

### 5.3.1. MMR Treatment with Partial Identification of the Population Outcome Distribution

I first focus on the identification problem generated by missing data and abstract from finite-sample imprecision in the data. To do this, I assume for now that the relevant observable features of the population are known. These are the distributions $P[y(t)|δ(t) = 1]$ and $P[δ(t)]$, $t \in \{a, b\}$. In particular, $e_t \equiv E[y(t)|δ(t) = 1]$ and $p_t \equiv P[δ(t) = 1]$ are known.

I assume that outcomes are bounded, with known finite lower and upper bounds $y_{0t}$ and $y_{1t}$, but otherwise assume no knowledge of the distributions of unobservable outcomes $P[y(t)|δ(t) = 0]$, $t \in \{a, b\}$. In this setting, Proposition 1 of Manski (2007b) shows that the treatment allocation that minimizes maximum regret over all possible distributions of missing data is

$$(12) \qquad z^*(b) = 1 \ \text{ if } \ (e_a - y_{1a})p_a + (y_{0b} - e_b)p_b + (y_{1a} - y_{0b}) < 0,$$

$$= 0 \ \text{ if } \ (e_b - y_{1b})p_b + (y_{0a} - e_a)p_a + (y_{1b} - y_{0a}) < 0,$$

$$= \frac{(e_b - y_{1b})p_b + (y_{0a} - e_a)p_a + (y_{1b} - y_{0a})}{(y_{0b} - y_{1b})p_b + (y_{0a} - y_{1a})p_a + (y_{1b} - y_{0b}) + (y_{1a} - y_{0a})} \ \text{ otherwise.}$$

Proposition 1 shows that the minimax-regret rule assigns all persons to treatment b if and only if this treatment dominates the alternative; that is, if and only if b yields higher mean welfare than a under all possible distributions of the missing data. It is easy to verify that b dominates a if and only if $(e_a - y_{1a})p_a +$ $(y_{0b} - e_b)p_b + (y_{1a} - y_{0b}) < 0$. Similarly, the minimax-regret rule assigns all persons to treatment a if and only if that treatment dominates b. When neither treatment dominates, the minimax-regret rule randomly assigns positive fractions of observationally identical persons to both treatments. The fraction allocated to treatment b increases with $e_b$, which measures the observable success of treatment b, and decreases with $e_a$, which measures the observable success of treatment a.

Suppose that $y_0 \equiv y_{0a} = y_{0b}$, $y_1 \equiv y_{1a} = y_{1b}$, and $p \equiv p_a + p_b \leq 1$. Inspection of (12) shows that neither treatment dominates when these conditions hold, whatever values $e_a$ and $e_b$ may take. The minimax-regret rule is the fractional treatment allocation

$$(13) \qquad z^*(b) = \frac{e_b p_b + y_1(1 - p_b) - e_a p_a - y_0(1 - p_a)}{(y_1 - y_0)(2 - p)}.$$

I call p the *observability score* for persons and I say that missing data are *prevalent* if $p \leq 1$. Missing data are necessarily prevalent when the data are from an observational study where treatments in the study population were selected in an unknown way. In this situation, $p_a$ and $p_b$ can be no larger than the probabilities that treatments a and b are chosen; hence, $p \leq 1$. In such a study, $p = 1$ if all realized outcomes are observed.

### 5.3.2. As-If MMR Treatment with Sample Data

Suppose now that the data are a random sample of the study population. Then the quantities ($e_a$, $e_b$, $p_a$, $p_b$) needed to compute the MMR allocation (12) are not known, but they may be estimated by their sample analogs. Let N denote sample size and let $[(e_{Nt}, p_{Nt}), t \in \{a, b\}]$ be the sample estimates. The as-if MMR decision using these estimates is the treatment allocation

$$(14) \qquad z_N(b) = 1 \ \text{if} \ (e_{Na} - y_{1a})p_{Na} + (y_{0b} - e_{Nb})p_{Nb} + (y_{1a} - y_{0b}) \ < \ 0,$$

$$= 0 \text{ if } (e_{Nb} - y_{1b})p_{Nb} + (y_{0a} - e_{Na})p_{Na} + (y_{1b} - y_{0a}) < 0,$$

$$= \frac{(e_{Nb} - y_{1b})p_{Nb} + (y_{0a} - e_{Na})p_{Na} + (y_{1b} - y_{0a})}{(y_{0b} - y_{1b})p_{Nb} + (y_{0a} - y_{1a})p_{Na} + (y_{1b} - y_{0b}) + (y_{1a} - y_{0a})} \text{ otherwise.}$$

In general, Manski (2007b) was only able to motivate allocation (13) asymptotically, pointing out that $\lim_{N \to \infty} [(e_{Nt}, p_{Nt}), t \in \{a, b\} = [(e_t, p_t), t, \in \{a, b\}]$ a. s. by the strong law of large numbers. However, a finite-sample result was obtained when $y_0 \equiv y_{0a} = y_{0b}$, $y_1 \equiv y_{1a} = y_{1b}$, and the value of p is known, with $p \leq 1$. Then (12) reduces to

$$(15) \qquad z^*(b) = \frac{e_b p_b + y_1(1 - p_b) - e_a p_a - y_0(1 - p_a)}{(y_1 - y_0)(2 - p)},$$

whose finite-sample analog is

$$(16) \qquad z_N(b) = \frac{e_{Nb} p_{Nb} + y_1(1 - p_{Nb}) - e_{Na} p_{Na} - y_0(1 - p_{Na})}{(y_1 - y_0)(2 - p)}.$$

Proposition 2 shows that (17) is a finite-sample MMR treatment allocation. The proof rests on two facts. First, expected welfare is a linear function of the treatment allocation. Second $E[z_N(b)] = z^*(b)$ when $y_0 \equiv y_{0a} = y_{0b}$, $y_1 \equiv y_{1a} = y_{1b}$, and $p \leq 1$.

Proposition 2 shows that $z_N(b)$ is a finite-sample MMR decision, not the unique such decision. Any unbiased estimate of the MMR decision with knowledge of observable features of the population is finite-sample MMR. A curious implication is that, under the assumptions of Proposition 2, there is no advantage to large sample size from the finite-sample MMR perspective. Observing one randomly drawn person receiving each treatment is as good as observing all such persons. I am not aware of other decision problems in which there exist unbiased estimates of the MMR decision with knowledge of observable features of the population.

## 5.4. As-If Decisions Using Confidence Sets as Set Estimates

In Section 3, I cautioned against use of classical hypothesis tests to make decisions. It would be better for planners making binary choices to use tests with good decision-theoretic properties. I suggested evaluation of performance by maximum regret, as illustrated in Sections 3.4 and 3.5.

Complementary to classical testing has been the practice of reporting confidence sets, these being set estimates of the true state that have specified coverage probabilities. A confidence set with uniform coverage probability $\alpha$ is a set estimate $S(\cdot)$ such that $Q_s[\psi: s \in S(\psi)] \geq \alpha$ for all $s \in S$. Coverage probabilities are well-defined whatever the nature of the state space. Although statistical theory long focused on settings where states are point-identified parameters of probability distributions, research in econometrics has shown that confidence sets with desired asymptotic properties remain well-defined and computable when states are partially identified.

Researchers have generally used confidence sets to quantify statistically imprecise inference, without reference to decision problems. Nevertheless, they could be used to make as-if decisions with criteria (1′) through (3′). An open question for future research is to determine the welfare consequences when these criteria are applied with confidence sets.

When addressing this question, I expect that it will be productive to abandon the conventional practice of a priori specifying the coverage probability of a confidence set. This practice mirrors classical testing, which restricts attention to test statistics that yield at most a specified probability of Type I error.

6. Conclusion

Inference makes no explicit appearance in Wald's statistical decision theory. Neyman (1962) remarked (p. 16): "In my opinion, the inferential theory solves no problem but, given a degree of earnestness of the authors, may create the illusion that it does." Nevertheless, inference-based SDFs may have desirable theoretic properties and be tractable. When this is so, inference-based SDFs are worthy of study. This paper has given illustrations in the domains of point prediction and treatment choice.

I remarked at the outset that another reason to study inference-based SDFs may be necessity, due to institutional separation between research and decision making. When planners observe inferences drawn by researchers rather than raw data, they can perform the mapping [inference → decision] but they cannot perform the more basic mapping [data → decision]. They can only choose among the inference-based SDFs made possible by research reporting practices. Given this, an important subject for future study is to learn the properties of SDFs that use the inferences provided by common reporting practices.